\documentclass[a4paper,11pt]{article}
\usepackage{jheppub} 
\usepackage{lineno}

\usepackage[english]{babel}

\usepackage{amsmath}
\usepackage{graphicx}

\title{Blind unblinding procedure for the PADME X17 data sample}

\author[a]{S. Bertelli,}
\author[a]{F. Bossi,}
\author[a]{R. De Sangro,}
\author[a]{C. Di Giulio,}
\author[a]{E. Di Meco,}
\author[a]{D. Domenici,}
\author[a]{G. Finocchiaro,}
\author[a]{L.G. Foggetta,}
\author[a]{M. Garattini,}
\author[a]{P. Gianotti,}
\author[a]{M. Mancini,}
\author[a]{I. Sarra,}
\author[a,1]{T. Spadaro%
\note{Corresponding author.},}
\author[a]{E. Spiriti,}
\author[a]{E. Vilucchi,}
\author[a,b]{V. Kozhuharov,}
\author[b]{K. Dimitrova,}
\author[b]{S. Ivanov,}
\author[b]{Sv. Ivanov,}
\author[b]{R. Simeonov,}
\author[c]{F. Ferrarotto,}
\author[c]{E. Leonardi,}
\author[c]{P. Valente,}
\author[c]{A. Variola,}
\author[c,d]{E. Long,}
\author[c,d]{G.C. Organtini,}
\author[c,d]{M. Raggi,} 
\author[e]{A. Frankenthal,}
\author[a]{F. Arias-Arag\`on,}
\author[f]{L. Darm\'e,}
\author[g]{G. Grilli di Cortona,}
\author[a,h]{E. Nardi}

\affiliation[a]{Istituto Nazionale di Fisica Nucleare, Laboratori Nazionali di Frascati,\\ C.P. 13, 00044 Frascati, Italy}
\affiliation[b]{Faculty of Physics, University of Sofia ``St. Kl. Ohridski''\\ 5 J. Bourchier Blvd., 1164 Sofia, Bulgaria}
\affiliation[c]{Istituto Nazionale di Fisica Nucleare, Sezione di Roma,\\ p.le Aldo Moro 5, I-00185 Rome, Italy}

\affiliation[d]{Physics Department, ``Sapienza'' Universit\`a di Roma\\ p.le Aldo Moro 5 00185 Roma, Italy}
\affiliation[e]{Department of Physics and Astronomy , University of California, Irvine, \\ Irvine, CA 92697-4575, USA}

\affiliation[f]{Universit\'e Claude Bernard Lyon 1, CNRS/IN2P3, Institut de Physique des 2 Infinis de Lyon,\\ UMR 5822, F-69622, Villeurbanne, France}
\affiliation[g]{Istituto Nazionale di Fisica Nucleare, Laboratori Nazionali del Gran Sasso,\\ Assergi, 67100, L’Aquila (AQ), Italy}
\affiliation[h]{Laboratory of High Energy and Computational Physic, HEPC-NICPB,\\ R\"avala 10, 10143, Tallin, Estonia}

\emailAdd{tommaso.spadaro@lnf.infn.it}

\abstract{
The PADME experiment at the Frascati DA$\Phi$NE LINAC has performed a search for the hypothetical X17 particle, with a mass of around 17 MeV, by scanning the energy of a positron beam striking a fixed target. The X17 should be produced from the resulting $e^+e^-$ annihilation. Since the expected mass of this particle is only roughly known, data sidebands cannot be clearly defined.
Furthermore, the need to keep the analysis blind to potentially sizable signal contributions prevents a clear assessment even of the quality of the data sample in this search.
In light of these challenges, this paper presents an innovative strategy adopted by the PADME Collaboration to perform data quality checks without disclosing the X17 sample. Moreover, the procedure designed to eventually unblind the data is described, together with the statistical approach adopted to extract the limits on the coupling between the X17 and the Standard Model.
}

\keywords{Analysis and statistical methods}

\begin{document}
\maketitle
\flushbottom
\tableofcontents

\section{Introduction}
The PADME experiment at INFN's Laboratori Nazionali di Frascati is a 
positron-on-fixed-target experiment operating at a center-of-mass (CoM) energy range of 
$14 < \sqrt{s} < 23$~MeV~\cite{Raggi:2014zpa}. The positron beam is provided by the PADME LINAC~\cite{Valente:2017mnr}. 
Following the observed anomaly in the angular spectrum of internal pairs produced in the de-excitation of nuclear states by the ATOMKI Collaboration~\cite{Krasznahorkay:2015iga} and the postulated existence of a particle with mass $M_X$ around 17~MeV (the ``X17'' particle), 
the Collaboration has focused its efforts on an independent search for the X17.\footnote{
A recent search for X17 has been performed by the MEG-II Collaboration~\cite{MEGII:2024urz}. Their result is still compatible with the ATOMKI observation at the 1.5~$\sigma$ level (see also Ref.~\cite{Barducci:2025hpg}).
}
With this goal in mind, the cross sections for the processes $e^+e^- \to e^+e^-$ and $e^+e^- \to \gamma\gamma$ in the energy range $16.5<\sqrt{s}<17.5$~MeV were measured during Run III, in late 2022.
Under the X17 particle hypothesis, the $e^+e^-$ production rate is expected to be enhanced, depending on the particle's coupling with the electromagnetic current. Considering an X17 vector-coupling strength $g_{ve}$ with electrons and positrons,
\begin{equation}
    {\cal L}\supset g_{ve} X^\mu_{17}\overline{e}\gamma_\mu e,
\end{equation}
enhancements are anticipated in a few of the energy points explored in the scan, which correspond to the X17 mass~\cite{Darme:2022zfw}.\footnote{An axion-like particle physics case can be considered as well. However, for simplicity, the present paper only refers to a vector X17.} 

Blind analysis procedures are crucial in searches for new physics signals in many fields of particle physics, particularly in dark matter searches and in studies of ultra-rare processes. Typically, part of the data is masked or made ``blind'' to the researchers. Only after the consistency of the acquired data samples is validated against the expected background estimates, the masked data can be analyzed in full. This procedure is called ``unblinding''. The consistency check is often achieved by using data sidebands, regions that are close to the signal region but where the absence of any new physics signal can be safely assumed.

The plan outlined above was also the original one 
for the analysis of the PADME data set collected during Run III.
The proposed mass scan region in Ref.~\cite{Nardi:2018cxi} was considered large enough to allow for meaningful signal-free regions under the hypothesis that positrons annihilate against at-rest target electrons.
However, after the realization that the electron motion significantly broadens the CoM energy of the collisions~\cite{Arias-Aragon:2024qji} and consequently the distribution of the potential X17 enhancement over the collected data sample, this approach had to be abandoned. The uncertainty on the X17 mass reported by the ATOMKI Collaboration and the broadening of the X17 production enhancement caused by the atomic electron motion leave no significant regions in which contributions from X17 production can be safely excluded. 

In this paper, we illustrate the strategy adopted by the PADME Collaboration to overcome the challenge in evaluating the quality of the data sample in the X17 mass region while remaining blind to the existence of the X17 particle. We also describe the forthcoming unblinding procedure to be followed once all analysis elements are frozen in place.

\section{PADME Run III analysis concepts and data handling}

The analysis of the Run III data set aims to select two-body final states after positrons annihilate against electrons when striking an active diamond target of 100~$\mu$m nominal thickness. A new physics signal ($e^+e^-\to \mathrm{X17} \to e^+e^-$) is searched for on top of background contributions from Standard Model (SM) processes ($e^+e^-\to e^+e^-$ or $e^+e^-\to\gamma\gamma$) via a finely spaced beam energy scan.

Preliminary studies of the beam features and their impact on the data analysis are described in Ref.~\cite{Bertelli:2024ezd}. We summarize the findings here:
\begin{itemize}
\item The positron beam energy $E_\mathrm{beam}$ is determined from a magnetic selection along the beam line. A beam energy absolute uncertainty of up to 2~MeV was assessed, corresponding to an absolute uncertainty on the CoM energy $\sqrt{s}$ of 30~keV. The uncorrelated systematic error in each point of the energy scan induces a negligible uncertainty, corresponding to less than a few keV in the CoM energy. The beam energy spread is around 750~keV or better, which corresponds to a standard deviation of approximately 20~keV on the value of $\sqrt{s}$.
\item The final states are selected requiring two in-time energy clusters in the PADME electromagnetic calorimeter (ECal), with cluster energies and positions consistent with the kinematics of an assumed two-body system. 
The number of two-body final states per positron on target (PoT), $R_2$, is given by:
\begin{equation}
\label{eq:master}
R_2(s) = \frac{N_2(s)}{N_\mathrm{PoT}} =  \left(B(s) + \epsilon_\mathrm{sig}(s)S(s,M_X,g_{ve})
\right),
\end{equation}
where the number of PoT ($N_\mathrm{PoT}$) and the number of two-body events ($N_\mathrm{2}$) are separately measured for various values of $\sqrt{s}$. The expected signal yield $S$ per PoT for given values of the mass and coupling of the X17 particle is determined from theory~\cite{Arias-Aragon:2024qji}, 
and includes contributions from the beam energy spread. The number of expected SM background events per PoT, $B$, and the signal selection efficiency $\varepsilon_\mathrm{sig}$, are determined from Monte Carlo (MC) simulations. 
\end{itemize}

The selected observable in Eq.~(\ref{eq:master}) suffers from an 18\% higher background rate compared to events with  
$e^+ e^-$-only final states,
but it benefits from lower systematic uncertainties and it allows for the neglect of uncertainties associated with the particle identification efficiency. 

\subsection{The PADME Run III data set}

The PADME X17 data set, named Run III, was collected from October to December 2022. It was acquired by varying the positron beam energy with a total of 47 different CoM energies $\sqrt{s}$. The scanning process covers the entire CoM region identified by the ATOMKI collaboration as significant for observing the postulated X17 particle. The scan covers the beam energy range 265--300~MeV, corresponding to values of $\sqrt{s}$ between 16.4 and 17.5~MeV.

Two additional data sets were also collected, one at a beam energy of 402~MeV and the other at five values ranging from 205 to 211~MeV. They correspond to  $\sqrt{s}$ values of 20.28~MeV, and from 14.5 to 14.7~MeV, respectively. These out-of-resonance samples are immune to contributions from X17, and are used for consistency checks and background studies, but their statistical power is too limited and their energy too far away from the X17 mass to serve as useful signal sidebands.

Following the recommendations in Ref.~\cite{Darme:2022zfw}, we collected on average approximately $10^{10}$ PoT per energy point. 
Scan points within the X17 region of interest have been collected with beam energy steps of approximately 0.75~MeV, comparable to the beam energy spread.

\subsubsection{Analysis-level corrections} 
During the reconstruction of raw data, several corrections are applied to account for variations in the data-taking conditions. These are determined per ``run'', a continuous data-taking period lasting up to 8 hours. Typically, one energy point in the scan includes three or more runs. 
The corrections include the energy scale of the calorimeter, which varies with 
temperature, and
the beam spot position and width at the target and at the ECal, which vary with the beam optics. 

A MC simulation is run for each energy point to determine the expected variations in $B(s)$ and $\varepsilon_\mathrm{sig}(s)$ in Eq.~(\ref{eq:master}). The expected point-by-point variations are at the level of several percent, thus exceeding the statistical fluctuations. 

\subsection{Statistical treatment of the data}
Since the number of two-cluster events in the presence of X17 is given by
Eq.~(\ref{eq:master}) as a function of $\sqrt{s}$, 
the signal extraction and sensitivity estimation are 
based on the discrimination power between the quantities 
\begin{equation}
\label{eq:master-1}
 B(s) \times\left( 1 + \frac{\varepsilon_\mathrm{sig}(s)}{B(s)}\times S(s,M_X,g_{ve})\right),
\end{equation}
and
\begin{equation}
\label{eq:master-2}
 B(s).
\end{equation}

Two separate scenarios can be considered:
The observation of statistically significant excesses in the event yields due to the presence of signal,
or upper limit setting on the X17 coupling parameter for different values of $M_X$. 
The procedure described below focuses on the latter, by deriving limits on the coupling strength $g_{ve}$.
We have chosen to employ a modified frequentist method, 
known as CLs, following the technique described in Ref.~\cite{ATLAS:2011tau}, 
with a test statistic and $\chi^2$ defined according to Refs.~\cite{Junk:2005awa}~and~\cite{Junk:2006fye}.

Let us denote with $L(S+B)$ and $L(B)$ 
 the likelihood functions in the signal+background and background-only hypotheses. The signal+background likelihood function depends on 
the X17 mass and coupling constant $M_X$ and $g_{ve}$. 
In addition, 
the likelihoods depend on a set of nuisance parameters $\theta$. 
The expected number of events $R_\mathrm{exp}(s) = R_\mathrm{2,exp}(s; M_X,g_{ve}, \theta)$, 
given the X17 mass $M_X$, coupling $g_{ve}$, and set of 
nuisance parameters, is determined via MC simulation.  
For PADME, the number of observed counts $N_2(s)$ for each energy scan point is of the order of 40,000 and therefore a 
Gaussian probability for the observed ratio $R_2$ is assumed.
The likelihood function is defined as
\begin{equation}
    L(\mbox{data} | M_X, g_{ve}, \theta) =  \left(\prod_s \frac{1}{\sqrt{2\pi \sigma_{R(s)}^2}} e^{-\frac{(R_2(s) - R_\mathrm{exp}(s))^2}{2\sigma_{R(s)}^2 }}\right) \times P(\theta),
\end{equation}
where $\sigma_{R(s)}$ includes the statistical uncertainty on $N_2(s)$ and the uncorrelated systematic uncertainty on $N_\mathrm{PoT}$.  
$P(\theta)$ is the probability for the particular set of  nuisance parameters $\theta$ to be the correct one.

The set of nuisance parameters $\theta$ is given by:
\begin{itemize}
    \item $B(s)$: The number of background events per PoT for each scan point. It can be parametrized as a linear function of $\sqrt{s}$, as determined from MC simulations;
    \item $f_\mathrm{PoT}$: To account for a possible systematic error of the calibration on the number of positrons on target, this scale correction is introduced;    
    \item $\varepsilon_\mathrm{sig}(s)/B(s)$: Signal efficiency for each scan point normalized to the background per PoT. It can be parametrized as a linear function of $\sqrt{s}$, as determined from MC simulations;  
    \item Three parameters describing the shape of the signal yield as a function of the CoM energy, for a given X17 mass and coupling: the intrinsic width of the resonance, the beam-energy spread, and the number of signal events produced at resonance.
\end{itemize}
To evaluate $P(\theta)$, the correlations among the parameters $B(s)$ and $\varepsilon_\mathrm{sig}(s)/B(s)$ are taken into account by assuming multivariate normal distributions. All other nuisance parameters are treated as independent single-variable normal distributions.

For given values of $M_X$ and $g_{ve}$, the constructed test statistic is ``Tevatron-like''~\cite{ATLAS:2011tau}:
\begin{equation}
    Q(M_X,g_{ve}) = -2 ~ \mathrm{ln} \frac{L_\mathrm{max}(s+b)}{L_\mathrm{max}(b)} = -2 ~ \mathrm{ln} \frac{L(\mbox{data}|M_X, g_{ve},\hat{\theta}_{(M_X, g_{ve})})} {L(\mbox{data}|\hat{\theta})},
    \label{eq:statistic}
\end{equation}
where $\hat{\theta}_{(M_X, g_{ve})}$ is the set of nuisance parameters
that maximize the likelihood in Eq.~(\ref{eq:master-2}) for given values of $M_X$ and $g_{ve}$, 
and $\hat{\theta}$ is the set of parameters that maximize the likelihood for the background-only hypothesis, i.e., assuming zero signal strength. 

As noted in Ref.~\cite{Junk:2006fye}, the test statistic in Eq.~(\ref{eq:statistic}) 
is equivalent to the difference of the generalized chi-square distribution $\Tilde{\chi}^2$ 
under the signal+background vs. background-only
hypotheses,
\begin{equation}
    Q(M_X, g_{ve}) = \Tilde{\chi}^2_{s+b} - \Tilde{\chi}^2_{b} = \Delta\Tilde{\chi}^2,
\end{equation}
taking into account the profiling of the systematic uncertainties via the nuisance parameters. 

For given $M_X$ and $g_{ve}$, following the procedure described in Ref.~\cite{ATLAS:2011tau} and generating multiple toy MC samples of pseudo-experimental data,
the test statistic $Q^\mathrm{obs}(M_X,g_{ve})$ and two $p$-values ($p_s(M_X,g_{ve})$ and $p_b$)
are computed, corresponding to the probability 
for the actual observation of such data in the 
signal+background and in the background-only hypotheses,
\begin{equation}
    p_s(M_X,g_{ve}) = P( Q(M_X,g_{ve}) \geq Q^\mathrm{obs}(M_X,g_{ve})),
\end{equation}
from samples generated with fixed nuisance parameters $\theta = \hat{\theta}_{(M_X, g_{ve})}$  and
\begin{equation}
    1 - p_b = P( Q(M_X,g_{ve}) \geq Q^\mathrm{obs}(M_X,g_{ve})),
\end{equation}
from samples generated with fixed nuisance parameters $\theta = \hat{\theta}$.
Then the $CL_s(M_X,g_{ve})$ is the ratio of the two probabilities:
\begin{equation}
    CL_s(M_X,g_{ve}) = \frac{p_s(M_X,g_{ve})}{1 - p_b}.
\end{equation}
If $CL_s(M_X,g_{ve}) < \alpha$, then, 
for a given mass $M_X$, coupling constants higher than 
$g_{ve}$ are excluded with $(1-\alpha)$ confidence level.

\subsection{The X17 line shape}
The existence of X17 will be revealed as an excess of two-cluster event yields at a certain
value of $\sqrt{s} \approx M_X$ in the energy scan data set. 
In fact, this excess is 
not just at a single energy scan point, but 
actually spreads over an extended $\sqrt{s}$ region
because of several contributions:
\begin{itemize}
    \item Beam energy spread: During Run III, the energy spread $\delta E/E$ was maintained at the level of  $0.25\%$ with a fractional error of 20\%;
    \item Motion of the atomic electrons.
\end{itemize}

The natural width of the X17 resonance, expected to be in the range $10^{-4} < \Gamma_{X17} < 10^{-1}$~eV~\cite{Nardi:2018cxi}, 
is much smaller than the beam energy spread, and therefore its contribution is negligible.

The contribution of the electron motion in the diamond target was studied in detail 
in Ref.~\cite{Arias-Aragon:2024qji}. 
The momentum distribution of the electrons
in diamond was obtained with two independent approaches --- through the Roothan-Hartree-Fock (RHF) wave functions,
and based on the material Compton profile --- 
leading to consistent results. 
For the Run III conditions, 
the electron motion effect was shown to be 
significant, leading to the
broadening of the X17 line shape by a factor greater than 2 with respect to the assumption that the electrons are at rest. 
The uncertainty on the signal shape (signal peak location and width) in the presence of electron motion derives from the uncertainty of the Compton profile data used. The fractional error amounts to a few percent overall.

\subsection{MC estimate of the expected sensitivity}

To illustrate the expected sensitivity of the analysis, 
a series of virtual pseudo-experiments were 
generated. For each pseudo-event, the nuisance parameters were sampled from the 
expected central values and associated uncertainties. 
The true number of positrons per 
energy scan point was fixed to $N_{PoT}(s) = 10^{10}$.  The measured quantities (observables) were the number of selected events and the measured number of positrons per energy scan point. These quantities were sampled from their respective probability density functions both at the level of generation of the pseudo-events and at the level of simulation of the MC toys with fixed nuisance parameters.  For each scan point, a total uncertainty of 1\%, 0.6\%, and 0.4\% on $N_\mathrm{PoT}$, $B(s)$, and $\epsilon_\mathrm{sig}$ are assumed, respectively. The uncertainty on the common scale factor $f_\mathrm{PoT}$ is assumed to be 1\%.
The actual systematic uncertainties will be estimated before the unblinding procedure.

\begin{figure}[ht]
\centering
    \includegraphics[width=0.8\textwidth]{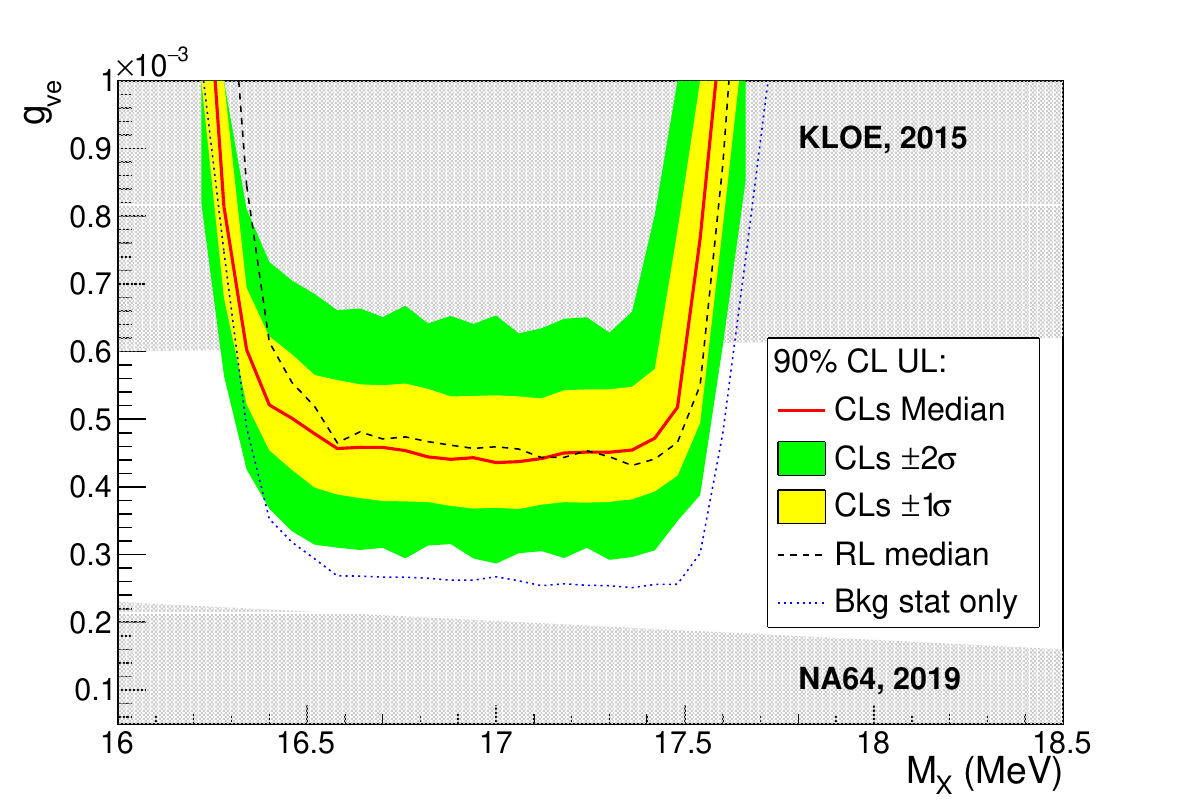}
    \caption{Expected 90\% confidence level upper limits in absence of an $X_{17}$ signal from the PADME Run III data sample, as a function of the $X_{17}$ coupling $g_{ve}$ and mass $M_X$. The median upper limit is shown in red. The $\pm1\sigma$ and $\pm2\sigma$ upper limit coverages are shown in yellow and green, respectively. The regions excluded by past searches from KLOE~\cite{Anastasi:2015qla} and NA64~\cite{NA642019} are shown in grey. The dashed line labelled ``RL median'' refers to the median upper limit from the log-likelihood ranked unified approach by Rolke and Lopez~\cite{Rolke:2000ij}, in presence of uncertainties of the expected background~\cite{Rolke:2004mj}.}
    \label{fig:MCsens}
\end{figure}

\begin{figure}[ht]
\centering
    \includegraphics[width=0.8\textwidth]{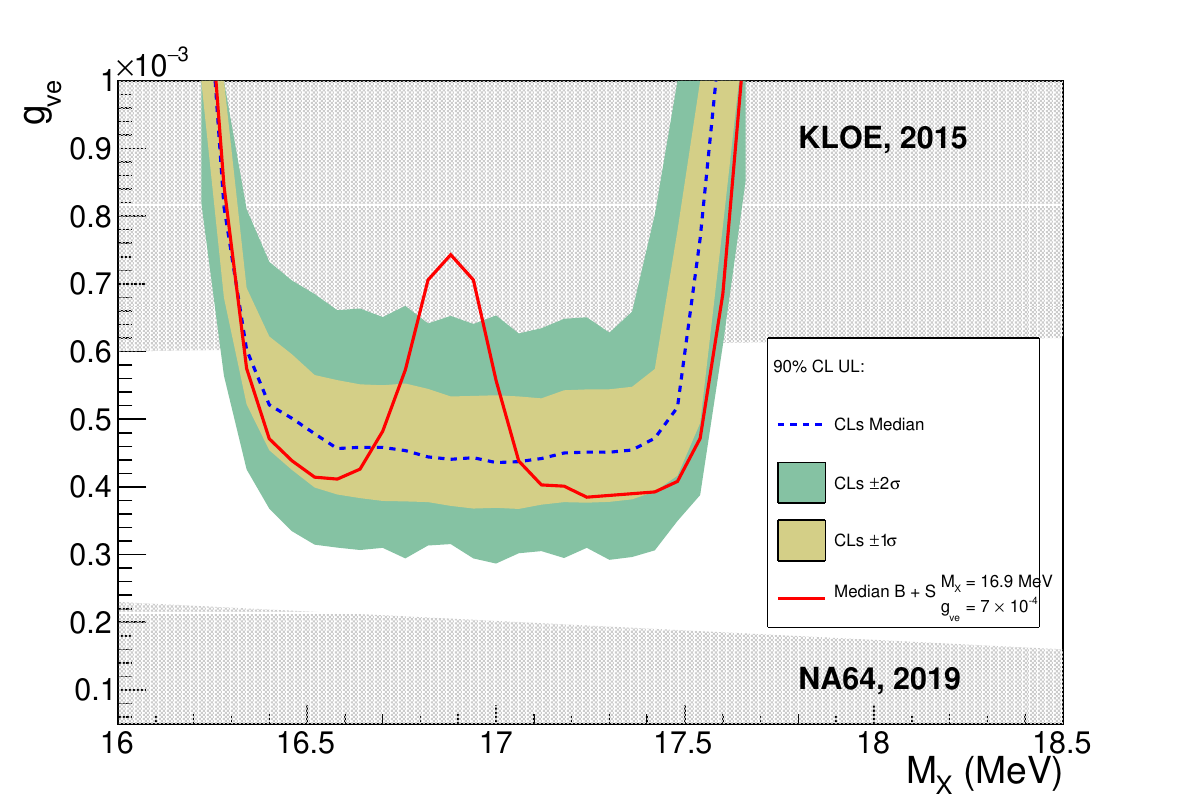}
    \caption{Expected 90\% confidence level upper limits in presence of an $X_{17}$ signal with $M_X = 16.9$~MeV and $g_{ve}=7\times10^{-4}$ from the PADME Run III data sample. The median upper limit is shown in red. The median upper limit in absence of signal is shown by the blue dashed line. The $\pm1\sigma$ and $\pm2\sigma$ upper limit coverages in absence of a signal are shown in dark yellow and dark green, respectively. The regions excluded by past searches from KLOE~\cite{Anastasi:2015qla} and NA64~\cite{NA642019} are shown in grey.}
    \label{fig:MCsensSig}
\end{figure}

In Fig.~\ref{fig:MCsens}, the expected 90\% confidence level exclusion limit in the absence of signal is shown. The red line is the median upper limit, while the yellow (green) bands represent the $\pm1\sigma$ ($\pm2\sigma$) quantiles. 
The median limit closely agrees with the median upper limit from the log-likelihood ranking method accounting for the sole background-related uncertainties~\cite{Rolke:2000ij,Rolke:2004mj}, represented by the dashed line labeled ``RL'' in Fig.~\ref{fig:MCsens}. The simulated samples that account for the full set of uncertainties from the nuisance parameters and from the observables lead to expected upper limits that are significantly weaker than those expected from pure background fluctuations, represented by the dotted blue line. The look-elsewhere effect was directly evaluated from the simulated samples and corresponds to a ratio of global to local probabilities of roughly 6; in absence of a signal, an observed upper limit exceeding $g_{ve}\approx 7\times10^{-4}$ for masses $M_X$ in the range 16.6--17.2~MeV corresponds to a probability below about 5\%. The median upper limit in the presence of a signal with $g_{ve}=7\times10^{-4}$ and $M_X = 16.92$~MeV is overlaid onto the signal-absent upper limit bounds in Fig.~\ref{fig:MCsensSig}.

The PADME Run III data set is expected to provide sensitivity to X17 masses and couplings in a region of parameter space still allowed by previous searches~\cite{Anastasi:2015qla,NA642019}. 

\section{Consistency of the data with the background-only hypothesis} 
\label{sec:gR}

The PADME data sample consists of 47 different values of the ratio $N_{2}/N_\mathrm{PoT}$, one for each energy scan point.
In the presence of signal, 
several points are affected.
However, no predefined 
sidebands free of signal exist to validate the
procedure, since the peak can be located anywhere 
in the scan region.   
The signal shape is wide and the data quality procedure must be blind to any signal contributions.

The main effects that impact the result extraction 
procedure are:

\begin{itemize}
\item The effect of radiative corrections: This has been estimated from Babayaga~\cite{Balossini:2006wc,Balossini:2008xr} MC runs by producing $e^+e^-(\gamma)$ and $\gamma\gamma(\gamma)$ final states. Radiative effects are expected to induce a linear variation in the ratio $N_2/(N_\mathrm{PoT}\times B)$ as a function of $\sqrt{s}$. The impact should be below 1--2\%.  

\item The absolute scale in the determination of $N_\mathrm{PoT}$: This is known with an uncertainty of up to a few percent, and is independent of $\sqrt{s}$. 

\end{itemize}

To circumvent the lack of a natural data sideband definition, an automatic procedure has been developed.  With this procedure, we are able to prove the consistency of data with the background-only expectation in a given sideband that is unknown to analyzers, and to determine best-fit parameters for the scan correction curve. These parameters might then fed back to the upper limit evaluation as additional nuisance parameters.

The size of the signal expected from theory 
drops to less than 10\% from its peak value for $E = E_\mathrm{res} \pm 6$~MeV~\cite{Arias-Aragon:2024qji}. For 
$g_{ve}\approx 8\times10^{-4}$, the signal yield for $E = E_\mathrm{res} \pm 6$~MeV~ corresponds to about 200 events, 
which is at the level of the statistical uncertainty 
of the number of background events. 
Therefore, any signal-induced excess is below a one-sigma background fluctuation for any scan point more than 6~MeV away from the scan resonance energy. In conclusion, at least 37 (31) energy scan points are unaffected by signal-induced effects at one (two) sigma level, provided that the coupling $g_{ve}$ is below $ 8\times10^{-4}$.

We define the ratio between the number of observed and expected events $g_R(s)$ as:

\begin{equation}
 g_R(s) = \frac{R_2(s)}{B(s)}.
 \label{eq:gR}
\end{equation}

If the positron flux and  the background efficiency were exactly determined, 
$g_R(s)$  would be around one in the absence of an X17 signal.
Given the uncertainty in the estimate 
of radiative effects, we assume that $g_R(s)$ is a  linear function of $\sqrt{s}$.

The following procedure aims to determine the location of 
the signal-free region and the linear bias $g_R(s)$, and to validate the systematic errors on $N_\mathrm{PoT}(s)$ and $B(s)$ established from MC and data-based studies, without unblinding the data set.

A linear fit is performed on  $g_R$ vs
$\sqrt{s}$
with a pre-determined number of continuous data points excluded from the fit, $N_s$, 
to account for a possible signal-induced bias. 
The start position of the 
masked region of consecutive $N_s$ points 
is chosen as the one that minimizes the
$\chi^2$ of the linear fit.

The outputs of the procedure are the best-fit parameters for $g_R(s)$ and the $\chi^2$ for the best fit. 
The exact location of the region excluded from the fit remains blinded throughout the procedure. 
The fit residuals are expected to be centered at zero, 
ideally with a standard deviation equal to the one from 
the individual points in the scan. 
From the fit $\chi^2$ 
and from the shape parameters of the pull distribution, 
the quality of the data set and the effectiveness 
of the applied corrections can be assessed. 

MC simulations in which a signal is injected show that even for the highest values of the X17 coupling considered: i) the statistically significant signal-affected bins (i.e., more than 2 standard deviations, given the acquired statistical power) are fewer than 10; ii) the region excluded by the fit is always centered around the hypothetical mass of the X17; and iii) the resulting fit parameters are unaffected by the presence of the X17 particle. Details are given in the following subsection.

\subsection{Application of the data consistency check to MC simulations}

\begin{figure}[ht]
    \centering    \includegraphics[width=0.45\textwidth]{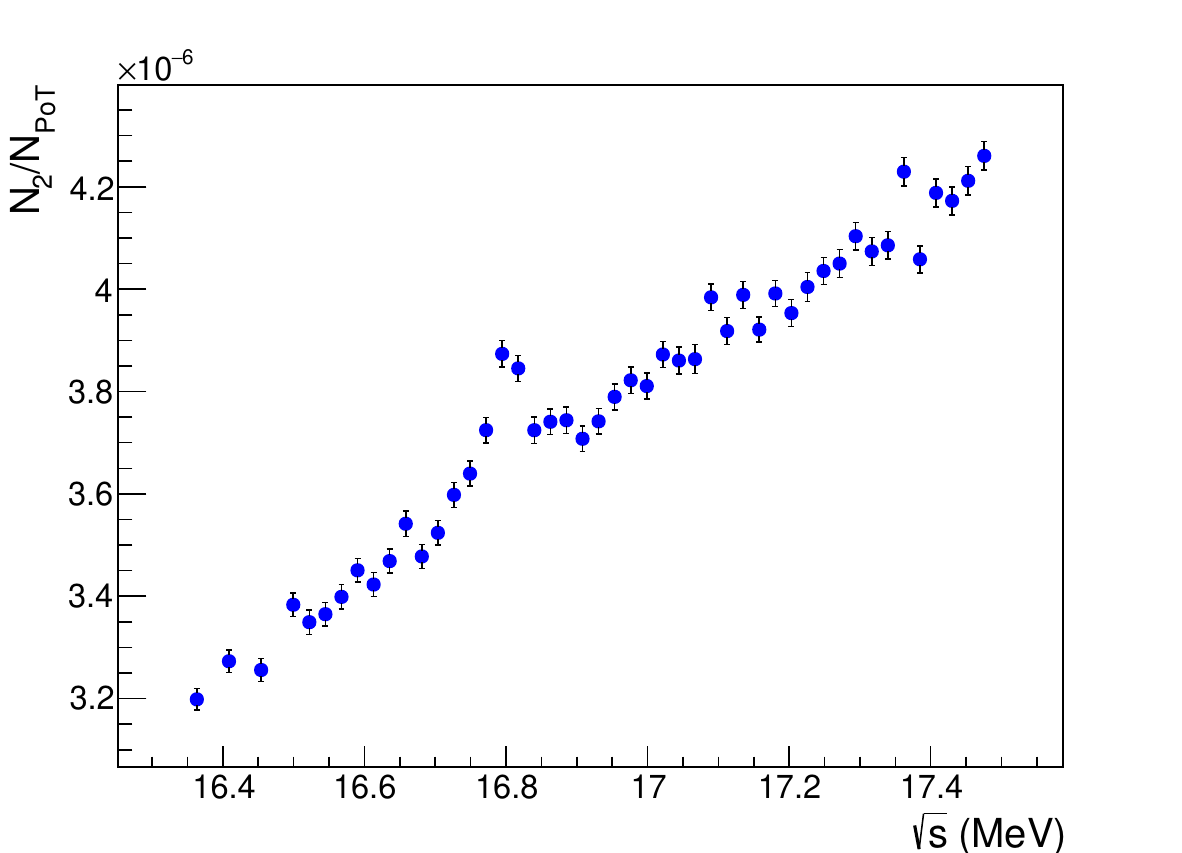}
\includegraphics[width=0.45\textwidth]{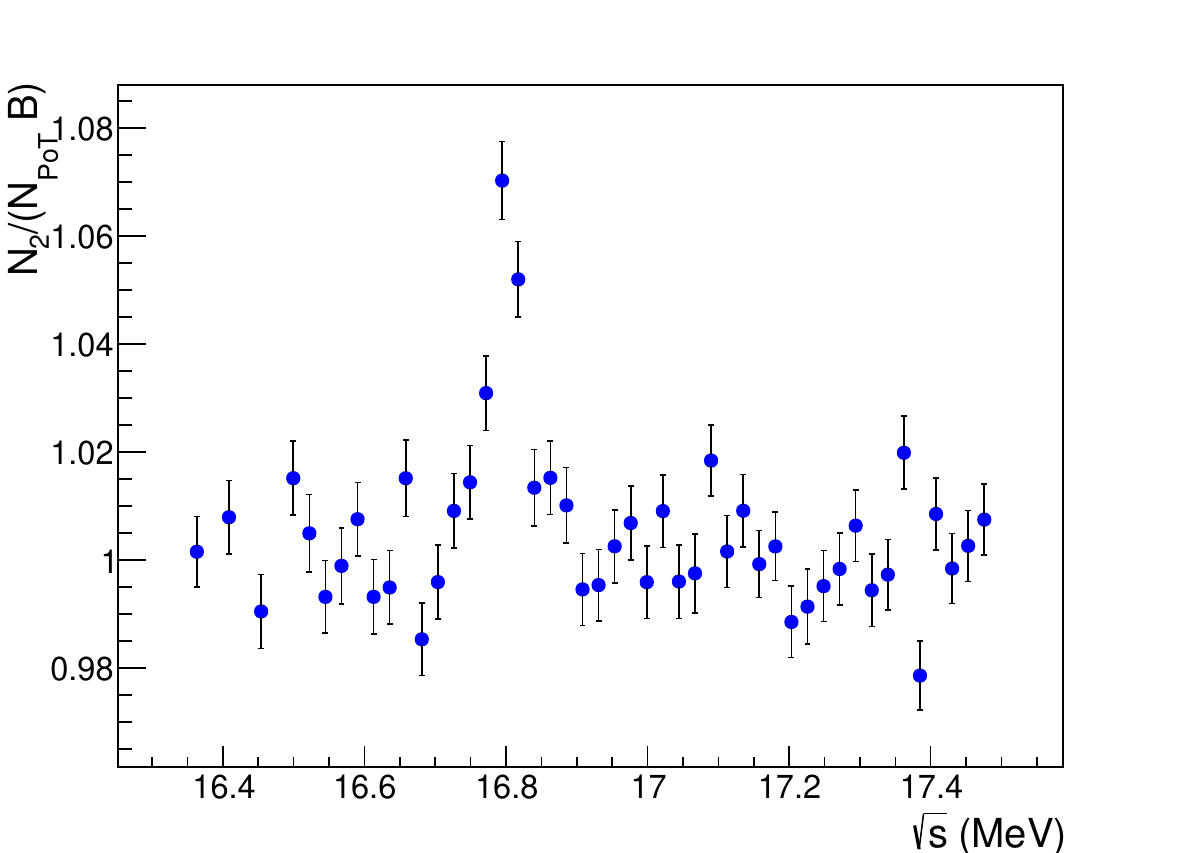}
    \caption{Left: Expected number of reconstructed two cluster events per positron on target for $M_{X17} = 16.8~\mathrm{MeV}$ and $g_{ve} = 7.9\times10^{-4}$. Right: $g_R$ as a function of $\sqrt{s}$.}
    \label{fig:RatioSignal}
\end{figure}

We used MC simulations to demonstrate that the data quality assessment procedure is blind to the existence of an X17 signal.
A complete data set including samples for all energy scan 
values collected in Run III was simulated, 
together with signal samples featuring several masses and coupling strengths.  
For each CoM value (i.e., energy scan point), the number of two-cluster events $N_2(s)$
divided by the number of positrons on target $N_\mathrm{PoT}$ was computed as a function of $\sqrt{s}$, as shown in Fig.~\ref{fig:RatioSignal}.

The rising slope of the uncorrected sample is 
dominated by the acceptance, which increases with beam momentum --- higher Lorentz boosts reduce the angle between outgoing particles. 
The acceptance correction obtained from the MC simulation also accounts for the cross section dependence on $\sqrt{s}$, resulting in a constant value of the estimated background.

\begin{figure}[ht]
    \centering
       \includegraphics[width = 0.45\textwidth]{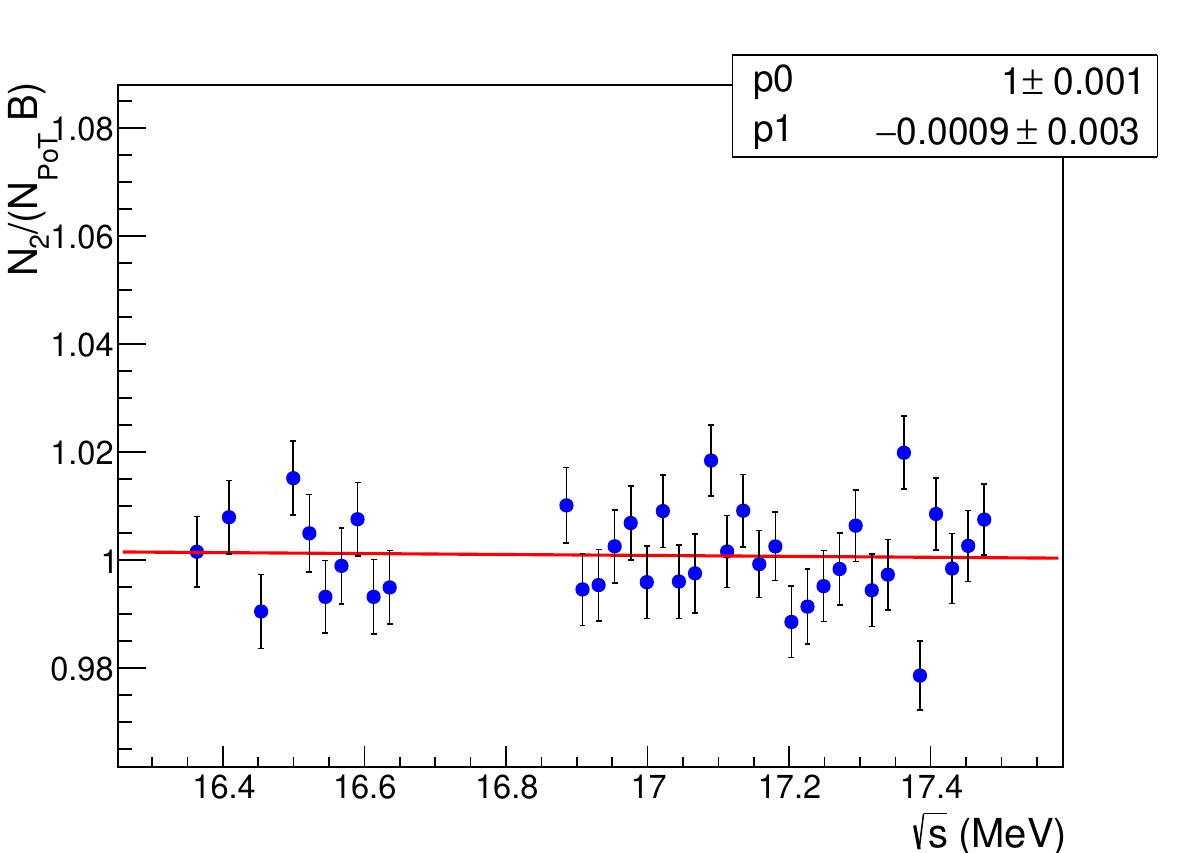}
        \includegraphics[width = 0.45\textwidth]{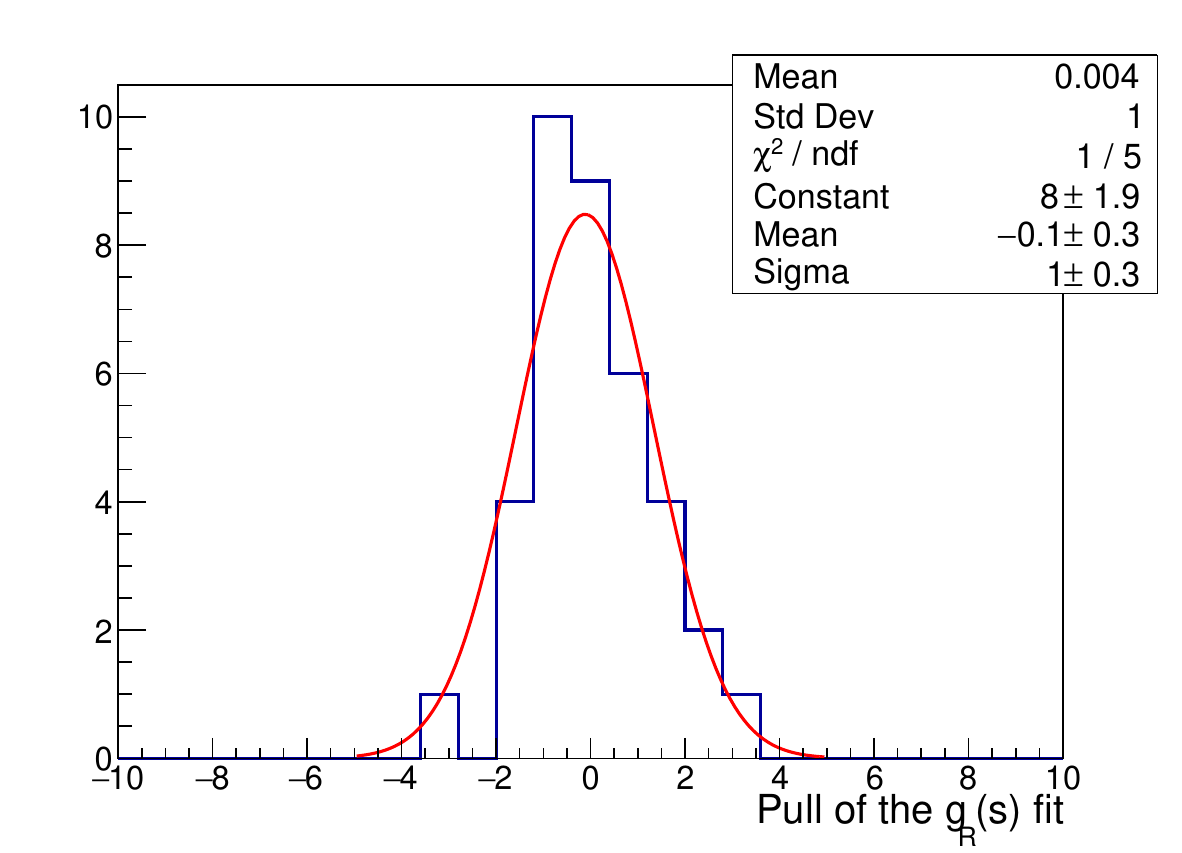}
    \caption{Left: Line fit to the remaining values of  $g_R(s)$  after masking the determined region to minimize the $\chi^2$ of the line fit using generated MC samples  for $M_X = 16.8~\mathrm{MeV}$ and $g_{ve} = 7.9\times10^{-4}$.  Right: Distribution of the pulls of the individual experimental points $g_R(s)$ with respect to the line fit.}
    \label{fig:RatioBlind}
\end{figure}

In Fig. \ref{fig:RatioBlind}, the dependence of $g_R$ on $\sqrt{s}$ 
is displayed after the blinding region has been identified by the automatic procedure. 
The procedure successfully localized the region to blind and restored the linearity.
The resulting $\chi^2$ of the fit is good and the pulls with respect to the linear fit can be safely used to assess data quality, since the signal is excluded.

\subsection{Validation of the procedure}
The developed methodology was extensively validated. MC events were produced with fixed values of the X17 parameters: the X17 mass was varied in the range $16.22 < M_{X} < 17.62$~MeV in steps of 20~keV (71 values in total) and the coupling was varied in the range 
$1\times 10^{-4} < g_{ve} < 7.9\times10^{-4}$, in steps of 0.35 (20 values in total).
In total, 1420 different and independent experimental outcomes were generated. Each outcome corresponds to the number of two-cluster events $N_2(s)$ for each 
energy scan point, 
the acquired statistics $N_{\mathrm{PoT}}(s)$,
the background yield per PoT $B(s)$, and the signal parameters. 
The values were obtained sampling each quantity independently according to its expected experimental uncertainty.

\begin{figure}[ht]
    \centering
    \includegraphics[width = 0.47 \textwidth]{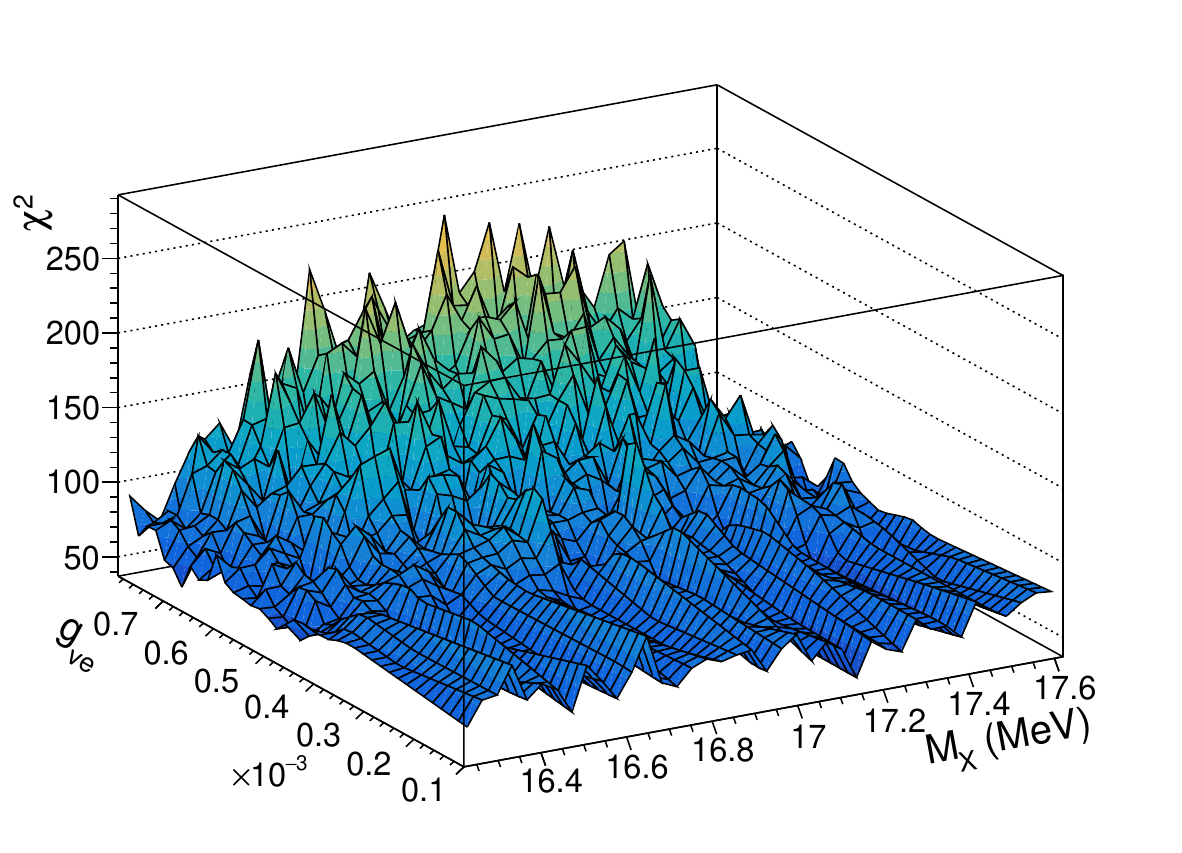}
    \includegraphics[width = 0.47 \textwidth]{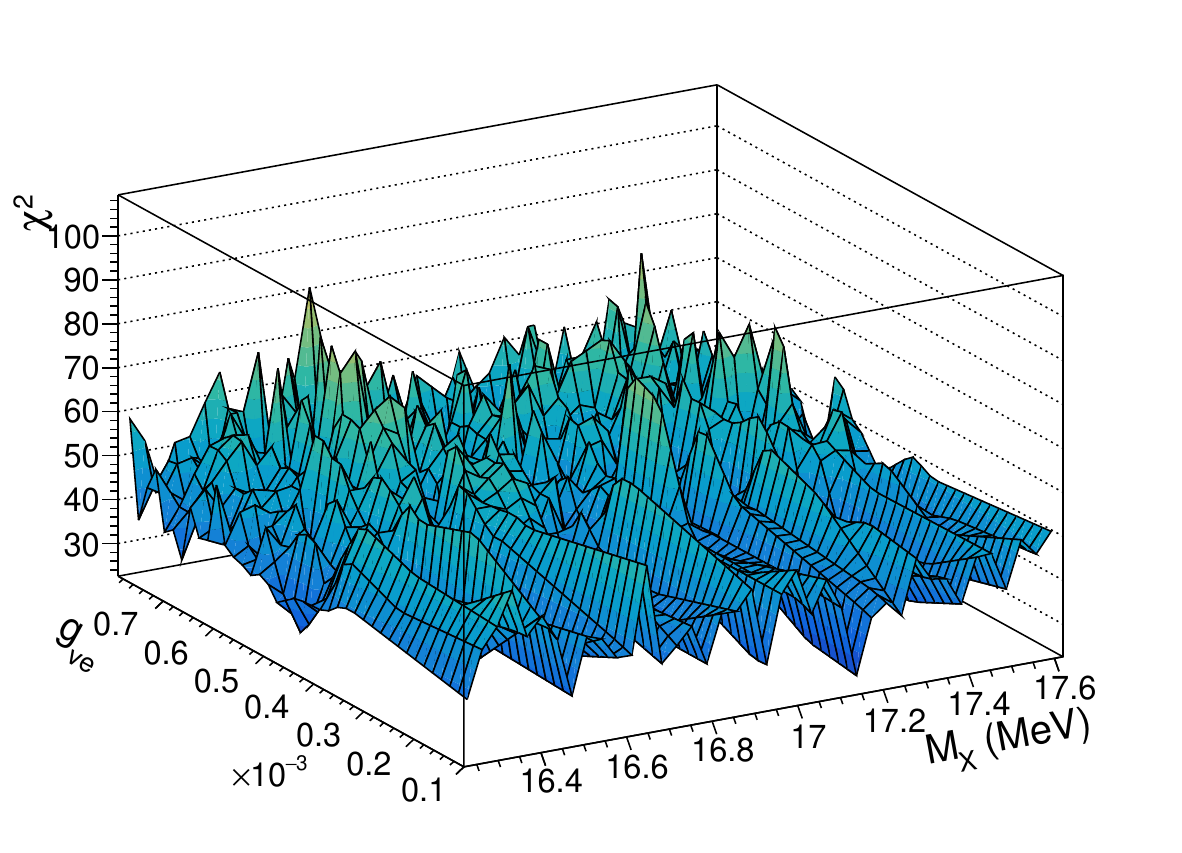}
    \caption{The distribution of the $\chi^2$ for a line fit 
    including all points (left) and masking a certain consecutive
    region to minimize the $\chi^2$ (right). }
    \label{fig:chi2-mask}
\end{figure}

A linear fit to $g_R(s)$ was performed for each of the virtual experiments
before employing the signal masking procedure. 
The resulting $\chi^2$ of the fits are shown on the left plot in 
Fig.~\ref{fig:chi2-mask}. The total number of scan points in each virtual experiment was the same as in the Run III data, 47. 
The $\chi^2$ distribution as a function of $M_{X}$ and $g_{ve}$
shows a clear rise as $g_{ve}$ increases, because of the 
injection of a larger amount of X17 signal.
After masking the ``signal region'', as described above, the $\chi^2$ distribution is uniform vs. 
$M_{X}$ and $g_{ve}$, with no visible structures. Still, for $g_{ve} > 5\times10^{-4}$, fluctuations may lead to an elevated $\chi^2$ for some of the virtual experiments. 

\begin{figure}[ht]
\centering    \includegraphics[width=0.3\textwidth]{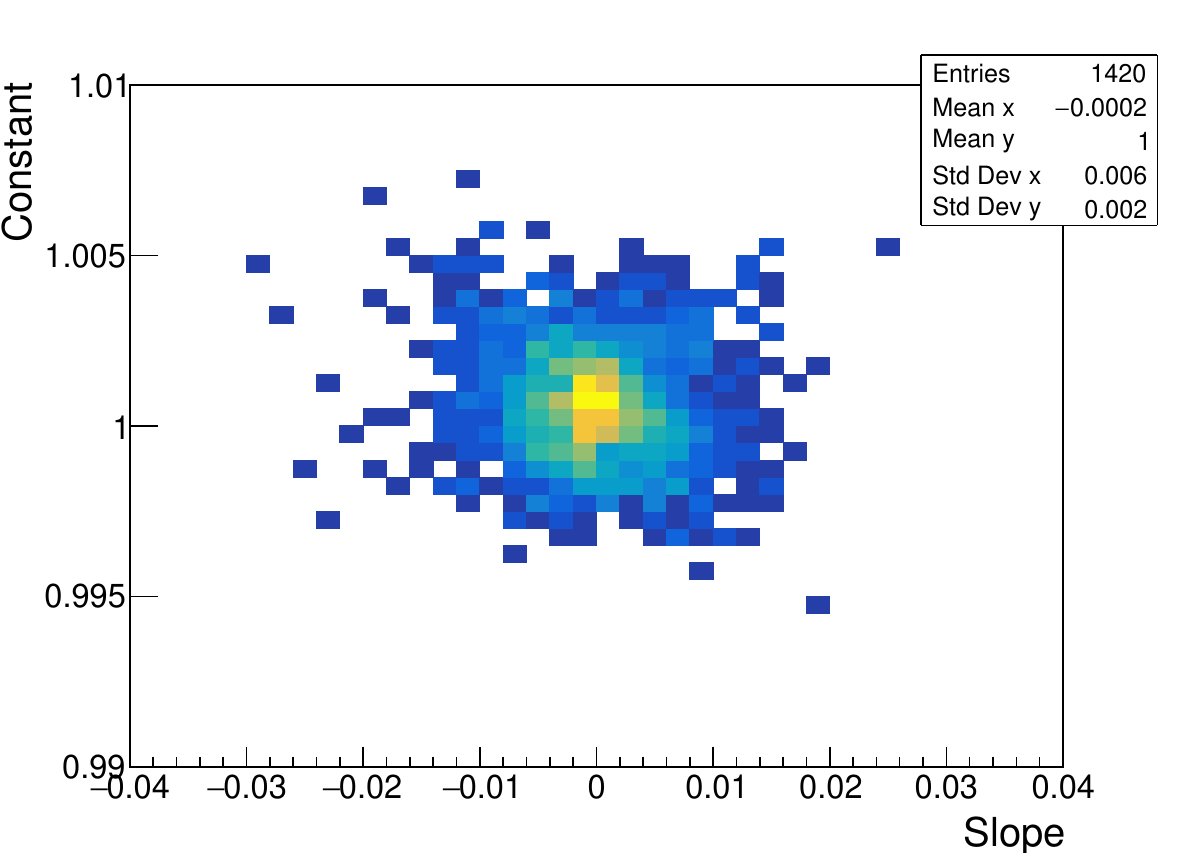}
\includegraphics[width=0.33\textwidth]{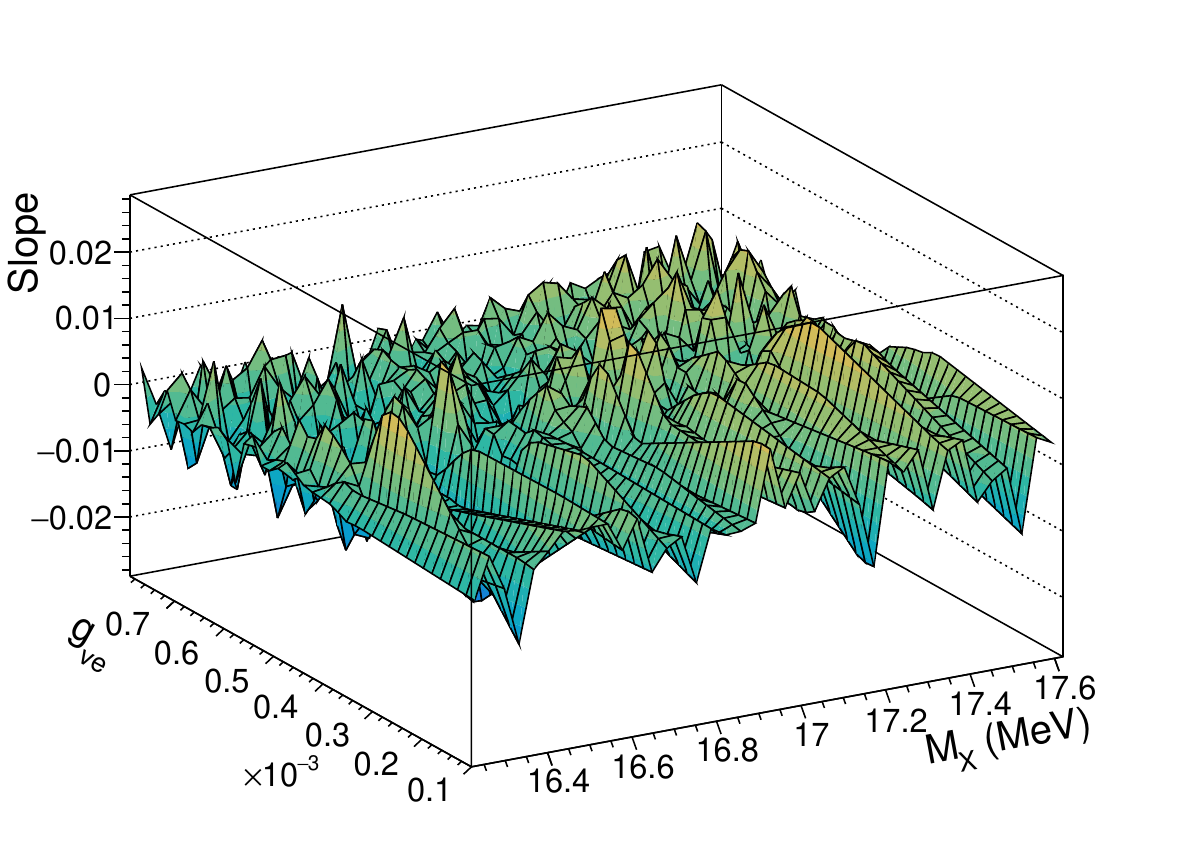}
\includegraphics[width=0.33\textwidth]{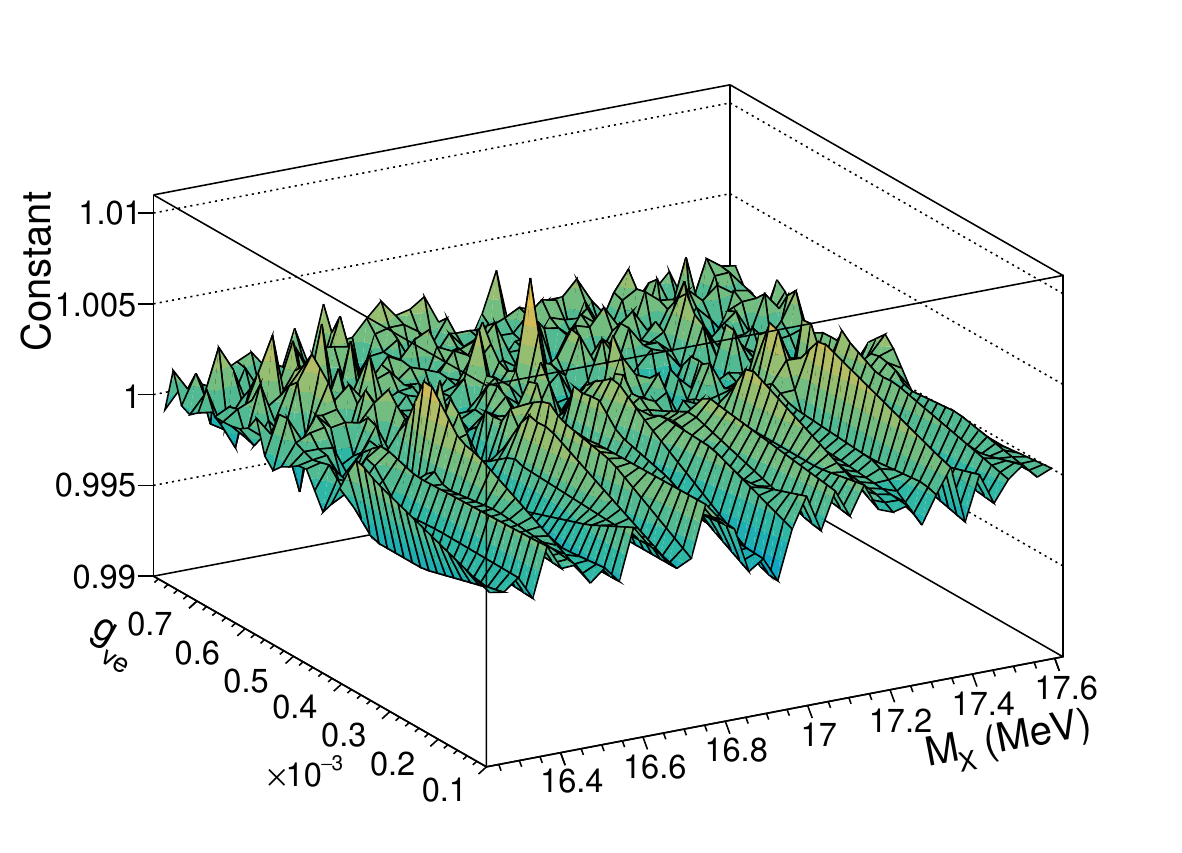}
\caption{Obtained values for the slope and the constant parameters from the linear fit performed after the identification of the masked region of $g_R$ as a function of $\sqrt{s}$ and their dependence on the X17 parameters - mass $M_{X17}$ and coupling constant $g_{ve}$.}
\label{fig:linear-fit-params}
\end{figure}

Since all virtual experiments were sampled from a distribution
with the mean $N_2(s)$ equal to the expected number of two-cluster events for a given $N_{PoT}$, 
the expected values for the constant and slope parameters of $g_R$ as a function of $\sqrt{s}$ are 1 and 0, respectively.
This is only true when the masked region successfully overlaps
with the ``signal region'', 
since for an $M_{X}$ close to 16.22~MeV the excess of events 
might push the slope towards negative values, while
for $M_{X}$ approaching 17.62~MeV the slope might be pushed towards positive values.
As can be seen from Fig.~\ref{fig:linear-fit-params}, 
the nominal values for the constant and the slope parameters are recovered 
in the masked samples, with no residual correlation between the two parameters. 
Moreover, the values of the parameters do not depend on the true values of the signal coupling strength and mass.

\begin{figure}[ht]
\centering    \includegraphics[width=0.4\textwidth]{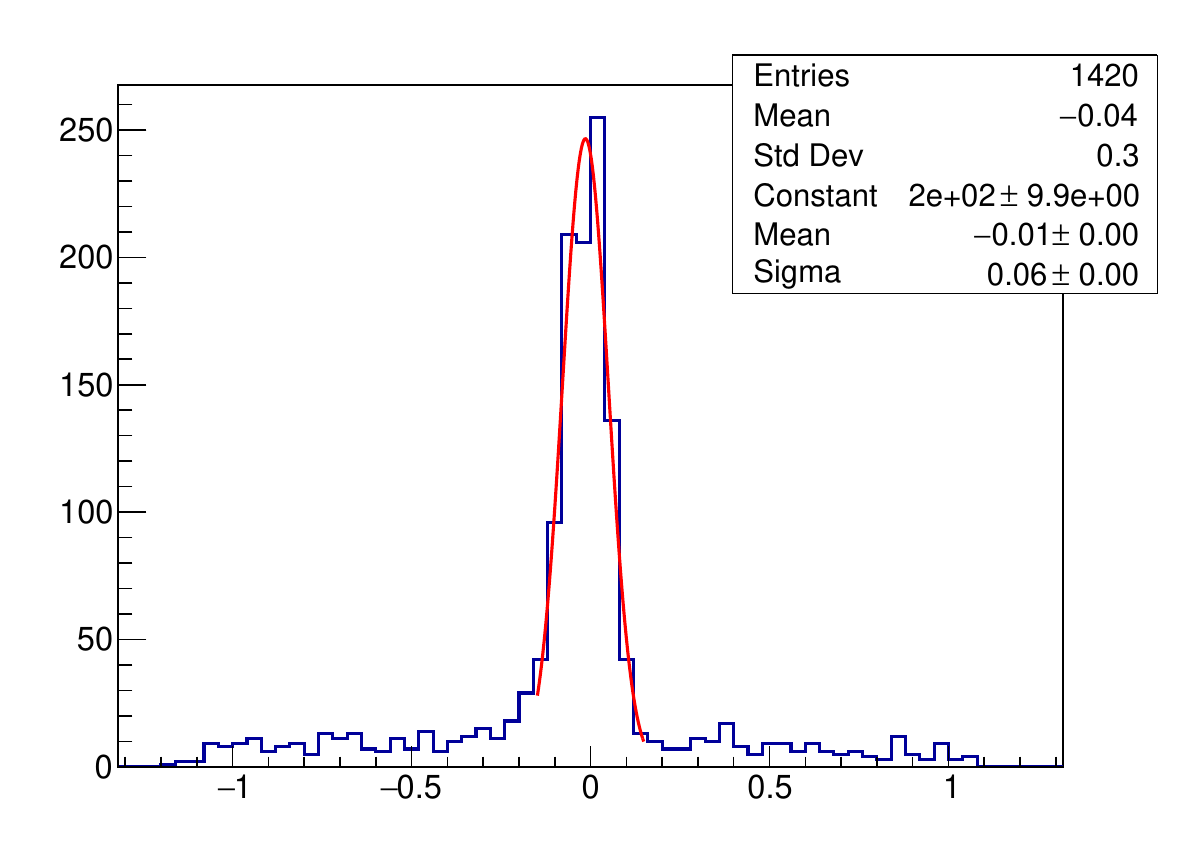}
\includegraphics[width=0.5\textwidth]{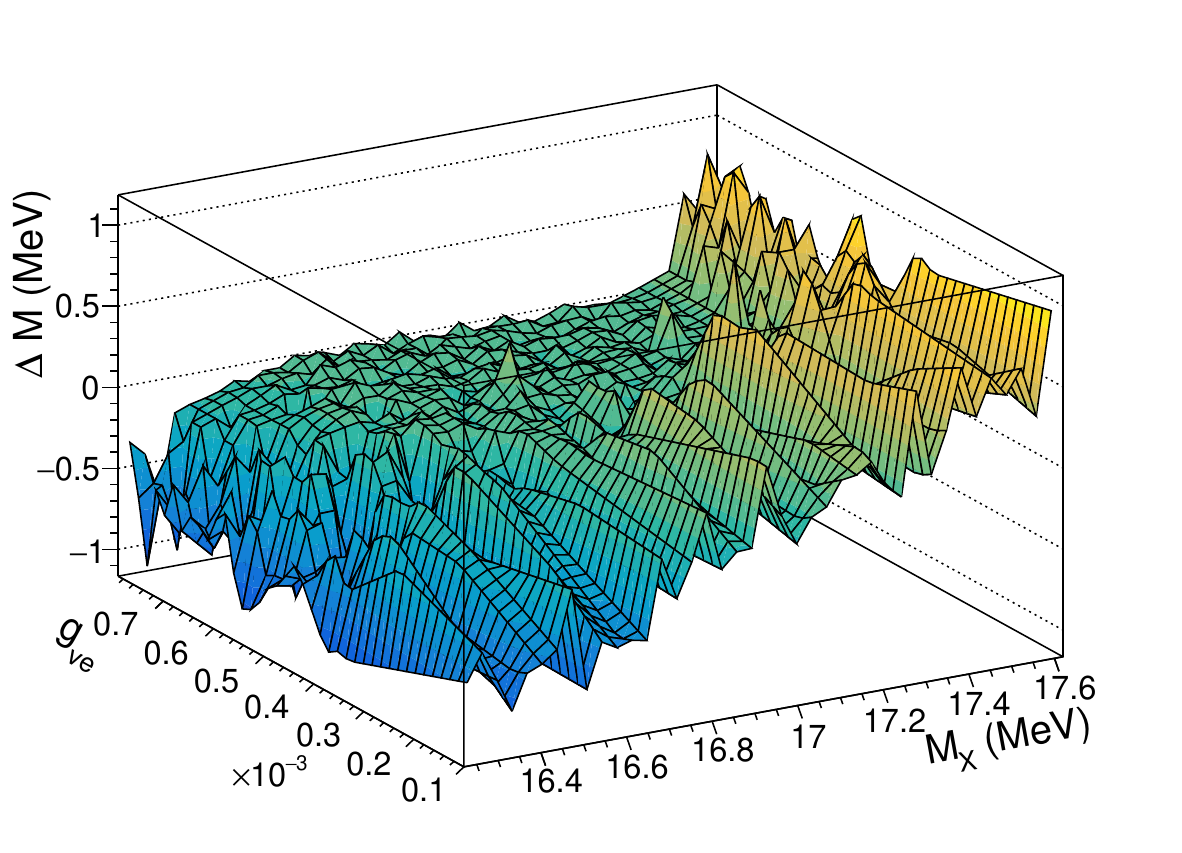}
\caption{Difference $\Delta M$ between the generated X17 mass and the central value of the masked region (left) and dependence of $\Delta M$ on the X17 mass and coupling constant (right).}
    \label{fig:mass-masked}
\end{figure}

The central mass value $M_{\mathrm{masked}}$ of the masked region provides an indication of the possible mass of the 
X17 particle. The difference $\Delta M$ between the generation value
$M_{X}$ and $M_{\mathrm{masked}}$ provides an indication of the 
successful identification of the signal region.
The $\Delta M$ distribution for all 1420 virtual experiments
is shown in the left panel of Fig.~\ref{fig:mass-masked}. 
While the intention of the 
procedure is not to reconstruct the X17 mass, the masking procedure clearly
successfully identifies the 
position of the X17 peak in most cases,
with an $M_{X}$ resolution on the order of 70~keV. 
This resolution, however, depends on the energy scan spacing and should not be taken as an indication of the 
strength of the method.

The tails in the $\Delta M$ distribution are associated with two effects. For low values of $g_{ve}$,
the signal contribution from the X17 is consistent with the statistical uncertainty of the background samples themselves,
and the masked region is randomly chosen within the energy scan range.
In addition, when $M_{X}$ approaches the 
borders of the scan interval, 
the masked region is either chosen at the beginning or at the end 
of the interval, with a fixed $M_{\mathrm{masked}}$ value
independent on the varied $M_{X}$, 
as can be seen from the right panel in Fig.~\ref{fig:mass-masked}.

The study of the data consistency procedure applied to a series of virtual experiments
demonstrates the lack of biases with respect to the possible existence of a signal and therefore ensures this is a robust method for assessing data quality.
\section{Unblinding procedure and extraction of results}
The analysis procedure discussed in this paper
is a robust method to perform 
blind data analysis while still allowing control of residual systematics and data consistency.
The  unblinding procedure consists of the following steps:
\begin{enumerate}
    \item Identify the 
    region to be masked and the related sideband region using a linear fit
    of $g_R(s)$ (Eq.~(\ref{eq:gR})). If the fit $\chi^2$ is good and the fit parameters are consistent with the MC simulation expectation within a few percent, proceed to the next step; 
    \item Unmask the fit pull distribution. If it is Gaussian, proceed to the next step;
    \item Unmask the data in the sideband region. If the plot of $g_R(s)$ vs. $\sqrt{s}$ does not show evident systematic dependencies, proceed to the next step;
    \item Unmask all the data and perform the statistical procedure to extract the observed upper limit. 
\end{enumerate}

\section{Conclusions}
This paper described the data-quality checks and unblinding procedures developed by the PADME Collaboration in the search for the X17 particle.
The multistep protocol enables an accurate assessment of the data quality in the signal sample and the validation of the expected systematic uncertainties without unblinding the analysis. A CLs method that will be used for the determination of the observed bounds on the X17 coupling strength and mass was also reported.

\section*{Acknowledgments}

The PADME collaboration acknowledges significant support from the Istituto Nazionale di Fisica Nucleare, and in particular the Accelerator Division and the LINAC and BTF teams of INFN Laboratori Nazionali di Frascati, for providing an excellent quality beam and full support during the data collection period. 
F.A.A., G.G.d.C. and E.N. are supported in part by the INFN ``Iniziativa Specifica'' Theoretical Astroparticle Physics (TAsP). F.A.A. received additional support from an INFN Cabibbo Fellowship, call 2022. The work of E.N. is supported by the Estonian Research Council, grant PRG1884. Partial support from the CoE grant TK202 ``Foundations of the Universe'', from the CERN and ESA Science Consortium of Estonia, grants RVTT3 and RVTT7, and from the COST (European Cooperation in Science and Technology) Action COSMIC WISPers CA21106 is also acknowledged.
Sofia University team acknowledges 
support by the European Union - NextGenerationEU, 
through the National Recovery and Resilience Plan of the Republic of Bulgaria project SUMMIT BG-RRP-2.004-0008-C01, 
by BNSF KP-06-COST/25 from 16.12.2024 based upon
work from COST Action COSMIC WISPers CA21106 
supported by COST (European Cooperation in Science and Technology), 
and 
by TA-LNF as part of STRONG-2020 EU Grant Agreement 824093.

\bibliographystyle{JHEP}


\bibliography{bibliography.bib}

\end{document}